\documentclass[pdflatex,iicol,sn-mathphys-num]{sn-jnl}

\usepackage{graphicx}%
\usepackage{multirow}%
\usepackage{amsmath,amssymb,amsfonts}%
\usepackage{amsthm}%
\usepackage{mathrsfs}%
\usepackage[title]{appendix}%
\usepackage{xcolor}%
\usepackage{textcomp}%
\usepackage{manyfoot}%
\usepackage{booktabs}%
\usepackage{algorithm}%
\usepackage{algorithmicx}%
\usepackage{algpseudocode}%
\usepackage{listings}%
\usepackage{graphicx}
\usepackage{subcaption}

\usepackage{array}         
\usepackage{caption}       

\usepackage{amssymb}  

\usepackage{rotating}
\usepackage{booktabs}
\usepackage{tabularx}
\usepackage{enumitem}
\usepackage{multirow}

\theoremstyle{thmstyleone}%

\theoremstyle{thmstyletwo}%

\theoremstyle{thmstylethree}%

\begin{document}

\title[Electrospinning-Data.org]{Electrospinning-Data.org, a platform for electrospinning data and analysis}

\author*[1]{\fnm{Mehrab} \sur{Mahdian}}
\email{mehrab.mahdian@taltech.ee}

\author[2]{\fnm{Ferenc} \sur{Ender}}
\email{ender.ferenc@vik.bme.hu}

\author[1,2]{\fnm{Tamas} \sur{Pardy}}
\email{tamas.pardy@taltech.ee}

\affil*[1]{
\orgdiv{Thomas Johann Seebeck Department of Electronics},
\orgname{Tallinn University of Technology}
}

\affil[2]{
\orgdiv{Department of Electron Devices},
\orgname{Budapest University of Technology and Economics}
}

\abstract{
Electrospinning produces nanofibers by depositing a polymer solution in an electrostatic field. Fiber morphology is governed by interacting material, solution, process, hardware, and environmental conditions, requiring extensive tuning. Yet observations remain dispersed across the literature in heterogeneous forms that hinder aggregation and reuse. We present Electrospinning-Data.org, the first open, community-oriented research infrastructure dedicated to electrospinning experimental data. Its domain-specific data model links materials, solution properties, process and hardware configurations, environmental conditions, and fiber outcomes, including unstable and unsuccessful experiments. Numeric quantities are harmonized to canonical units and morphology is represented through controlled descriptors. The platform supports guided web, batch, and programmatic contributions, governed by automated structural validation and expert moderation. Its continuously evolving data are complemented by immutable DOI-bearing releases that support reproducible analyses. Public web interfaces, a REST API, and a Python client enable interactive and programmatic access. The integrated Virtual Lab supports exploratory analysis, machine-learning modeling, model interpretation, and evidence-linked parameter exploration. Together, these components establish a failure-aware foundation for reproducible, data-centric electrospinning research.
}


\maketitle

\section{Introduction}

Open, structured, and reusable experimental data are increasingly recognized as foundational to reproducible and data-driven materials research \cite{scheffler_fair_2022}. The convergence of experimental materials science, FAIR \cite{wilkinson_fair_2016} (Findable, Accessible, Interoperable, and Reusable) data practices, and machine learning is enabling a data-centric, or "fourth paradigm" \cite{gray_jim_2009}, mode of scientific inquiry in which systematically curated datasets can reveal relationships and trends that are difficult to identify within individual studies \cite{draxl_nomad_2018}. Realizing this potential in an experimental domain requires more than the accumulation of isolated datasets: it requires an evolving research infrastructure capable of transforming heterogeneous experimental observations into coherent, queryable, and reusable records while preserving their provenance and supporting community contribution.

Electrospinning is a widely used method for producing polymer micro- and nanofibers whose morphology and properties emerge from the interaction of numerous experimental factors, including material and solution properties, process parameters, hardware configuration, and ambient conditions \cite{bhardwaj_electrospinning_2010, xue_electrospinning_2019, mqatywa_multiparametric_2026}. Because these variables jointly determine the outcome, individual studies typically sample only a small region of a large experimental parameter space \cite{guo_research_2022, mahdian_cogni-e-spindb_2026}. The broader scientific value of the accumulated literature therefore depends strongly on whether observations from different laboratories, materials, and experimental configurations can be systematically aggregated, compared, and reused. Electrospinning is consequently well suited to a data-centric approach, yet its experimental process-structure-property data remain difficult to reuse at scale.

Assembled into consistently structured records that link experimental inputs to observed outcomes, such data provide the empirical basis for data-driven modeling of process-structure-property relationships. Because the relationship between processing conditions and fiber morphology is generally nonlinear and material-dependent \cite{yordem_effects_2008}, these records enable statistical and machine-learning models that predict fiber characteristics from experimental conditions \cite{kalantary_mlr_2020}. These models can, in turn, inform experimental design and process optimization, reducing reliance on exhaustive trial-and-error, and can supply consistent reference data to complement physics-based simulation \cite{mahdian_surrogate-based_2026, roldan_fibrecastml_2026-1, roldan_spincastmlanopen_2026-1}. Aggregated data further supports cross-study analysis capable of quantifying the influence of individual factors, including ambient conditions such as temperature and humidity, whose effects are difficult to isolate within any single experiment \cite{de_vrieze_effect_2009, mahdian_cross-model_2026}. Together, these opportunities motivate treating electrospinning data as a shared, reusable scientific resource rather than merely as a by-product of individual studies.

Experimental results are dispersed across the primary literature and are commonly reported in narrative text, tables, figures, and supplementary materials using heterogeneous terminology, incomplete parameter descriptions, and inconsistent representations of fiber morphology \cite{mahdian_cogni-e-spindb_2026}. Such data are rarely available as structured, machine-readable records that explicitly connect experimental conditions with observed fiber outcomes while preserving source provenance \cite{subeshan_machine_2024, mahdian_cogni-e-spindb_2026}. Consequently, researchers conducting cross-study analyses must extract, digitize, standardize, and reconcile information from individual publications before computational analysis. 

Our previous systematic collection of the electrospinning literature illustrates the scale of this problem. An initial search identified 318 candidate sources, of which 146 contained sufficiently clear quantitative experimental information for extraction \cite{mahdian_towards_2025}. These sources yielded 1,913 preliminary experimental records, many of which contained missing parameters; subsequent harmonization, completeness filtering, duplicate removal, and data-quality criteria resulted in Cogni-e-SpinDB 1.0 \cite{mahdian_cogni-e-spin_2025, mahdian_cogni-e-spindb_2026} , comprising 809 complete records from 57 sources across 12 precursor materials. Environmental parameters were among the least consistently represented categories, despite the established influence of factors such as temperature and humidity on electrospinning outcomes. Such reporting heterogeneity constrains systematic comparison and meta-analysis, reduces the amount of literature-derived data that can be directly reused, and creates a substantial preprocessing burden for data-driven and machine-learning studies \cite{himanen_datadriven_2019, mahdian_surrogate-based_2026}.

Existing efforts to organize materials data address different parts of this challenge. Within electrospinning, the Electrospun Fiber Experimental Attributes Dataset (FEAD) \cite{sarma_electrospun_2023} provides curated experimental records centered on poly(vinylidene fluoride) (PVDF), while our previous Cogni-e-SpinDB 1.0 \cite{mahdian_cogni-e-spin_2025, mahdian_cogni-e-spindb_2026} expanded the scope to a machine-learning-ready dataset containing 809 experimental records across multiple polymer systems. Cogni-e-SpinDB was designed as a stable dataset release and explicitly identified a companion web platform as a mechanism for collecting validated community contributions that could subsequently feed future versioned releases. At a broader scale, materials-science infrastructures such as the Materials Project \cite{horton_accelerated_2025}, NOMAD \cite{draxl_nomad_2018}, the Materials Data Facility (MDF) \cite{blaiszik_materials_2016, blaiszik_data_2019}, and Materials Cloud demonstrate how structured data services can support persistent access, computational reuse, and integrated scientific workflows. Domain-oriented resources in adjacent fields provide complementary models: Catalysis-Hub.org \cite{winther_catalysis-huborg_2019} combines continuously ingested computational catalysis data with web and programmatic access, while the Data Archive for the BRAIN Initiative (DABI) \cite{duncan_data_2023} combines controlled sharing of heterogeneous experimental data with facilities for discovery, visualization, and analysis.

These resources establish important precedents, but their data models and scientific assumptions do not directly represent the experimental semantics of electrospinning. General materials infrastructures are primarily organized around other classes of materials information and are not designed to capture the complete relationship between electrospinning materials, solution properties, processing conditions, equipment configurations, ambient conditions, and resulting fiber morphology. Conversely, electrospinning-specific datasets provide valuable structured records but remain principally dataset releases rather than continuously evolving community infrastructures. A further challenge is the heterogeneous description of nanofiber morphology itself \cite{mahdian_towards_2025}. To address this, our associated morphology framework provides a hierarchical representation of fiber characteristics spanning structure, topography, shape, composition, size, size variation, and defects, providing a basis for more consistent morphological annotation and computational reuse.

These considerations suggest that an open research infrastructure for electrospinning should:

\begin{enumerate}
    \item represent experiments through a structured, domain-specific schema linking materials, solution properties, process parameters, hardware configurations, and environmental conditions to observed fiber outcomes, including controlled morphology descriptors, instability and failure information, associated images, and source and contributor provenance;

    \item support community contribution and subsequent revision of experimental records through a governed data lifecycle combining automated structural validation with expert moderation before public release;

    \item provide versioned, citable, FAIR-aligned access to curated data, distinguishing the continuously evolving dataset from stable and referenceable archival releases;

    \item expose programmatic interfaces for both data retrieval and authenticated contribution, allowing external software, instrumentation, and scripted scientific workflows to interoperate with the resource; and

    \item connect the curated data directly to an integrated analytical environment supporting working-dataset construction, exploratory data analysis, supervised machine-learning modeling, model interpretation, and evidence-based parameter exploration, while maintaining traceability between computational outputs and the underlying experimental observations.
    
\end{enumerate}

With these requirements in mind, we present \textbf{Electrospinning-Data.org}, a community-oriented research infrastructure for the structured collection, curation, dissemination, and computational reuse of electrospinning experimental data. The platform integrates a domain-specific experimental data model and controlled morphology representation with community contribution and expert moderation, versioned public dataset releases, programmatic read and write access, and the \textbf{Virtual Lab}, an integrated environment for exploratory analysis, machine-learning modeling, model interpretation, and evidence-linked parameter exploration. Together, these choices prioritize sustained growth, provenance, and reuse over the simplicity of a one-time dataset publication.

\section{Method}

\subsection{Overall Platform Architecture}
\begin{figure*}[t]
    \centering
    \includegraphics[width=\textwidth]{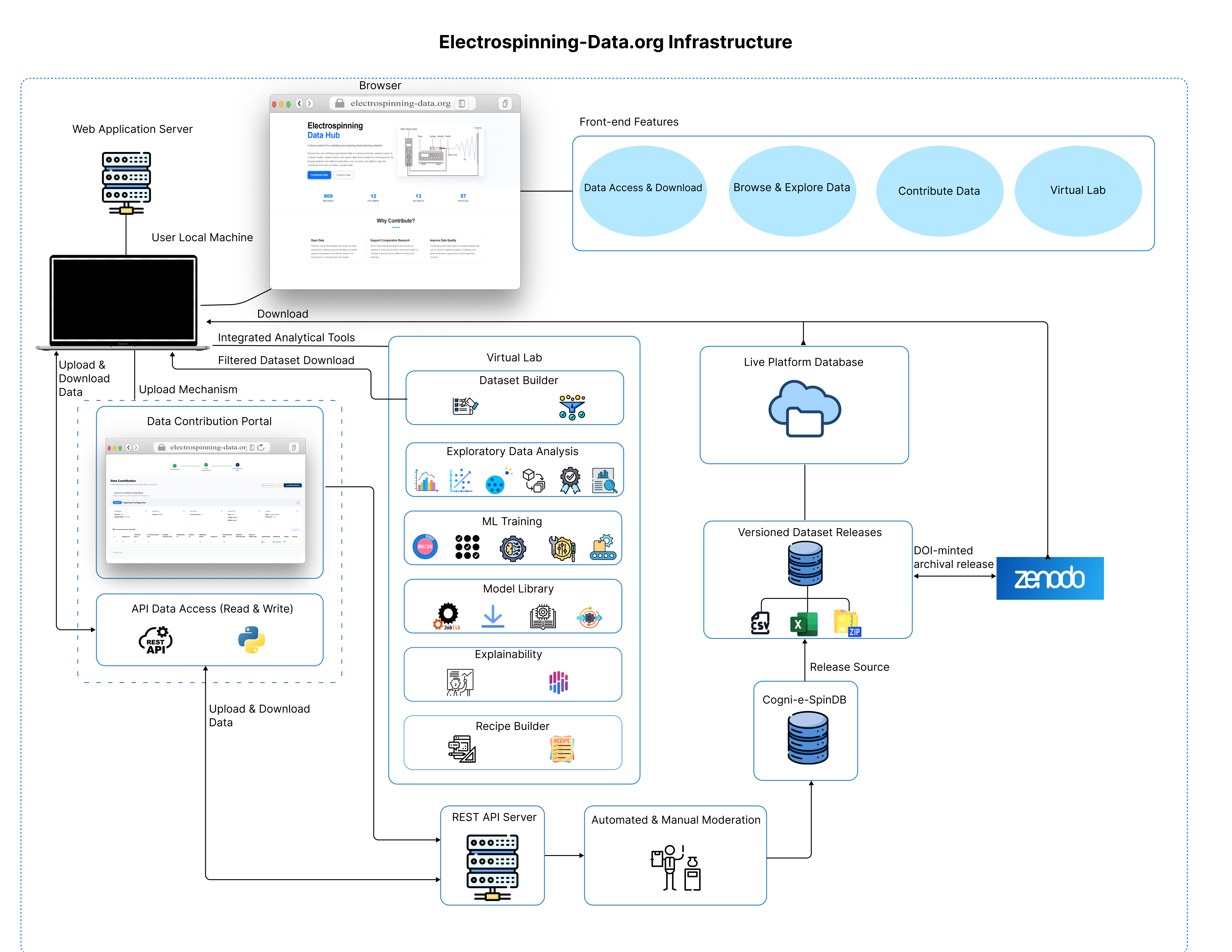}
    \caption{Architecture of Electrospinning-Data.org. Four cooperating tiers: a web client, a REST application backend, a persistent data store, and a separate Python computational engine, supporting the full cycle from contribution to reuse. Records enter through the web contribution portal or authenticated programmatic access, undergo automated validation and expert moderation, and once approved become publicly retrievable through the web interface, REST API, and Python client. The Virtual Lab couples the curated data to the computational engine's analytical modules, including exploratory data analysis, machine-learning training and interpretation, and the Recipe Builder. Periodic versioned snapshots are released on the platform and deposited to Zenodo with persistent DOIs.}
    \label{fig:platform-diagram}
\end{figure*}

Electrospinning-Data.org is designed to support the full cycle of electrospinning data aggregation and modeling, from community contribution and curation through discovery and computational analysis to reuse. The platform is organized as four cooperating tiers; a web client, a REST application backend, a persistent database, and a separate Python computational engine, that together expose the functionalities summarized in Fig. \ref{fig:platform-diagram}: (1) structured contribution of experimental records through a guided web submission workflow or authenticated programmatic write access; (2) automated structural validation followed by human moderation; (3) representation of each experiment in a common schema spanning material, solution, process, and environmental parameters, fiber outcome and morphology information, and associated source and contributor metadata; (4) public discovery and access through dataset browsing, filtering, and record preview, versioned dataset snapshots, and a REST read API supporting pagination, version selection, and downloadable representations in JSON, Excel, CSV, and image-archive formats; (5) an authenticated computational workspace (the Virtual Lab) that couples the curated dataset to reusable analysis and modeling services provided by the Python engine; (6) exploratory data analysis, supervised machine-learning training and evaluation, retention and reuse of trained models, and SHAP-based model interpretation; (7) evidence-based parameter exploration through the Recipe Builder, which derives candidate parameter configurations from matching experimental records and surfaces the supporting records and summary statistics; and (8) module-level and dataset-level export for downstream reuse and reporting, together with the administrative governance functions that sustain the resource. In addition, a Python client library (electrospinning-data-client), installable with pip, wraps the REST read and write endpoints into a single programmatic interface, exposing dataset download, search, and versioned retrieval alongside authenticated record submission and update. Technical detail for each tier is provided in the subsections that follow.

\subsection{Experimental Data Model and Database Architecture}
\label{sec:database}
Data are stored in a MySQL relational database on a dedicated institutional server with regular backups. Records are organized into related tables so that SQL selections return the subset satisfying given experimental criteria. Separate tables store provenance, experimental inputs, and outcomes, supporting one-to-many and many-to-many mappings such as multi-component polymer/solvent systems and multi-label morphology assignments. A central table anchors each experiment and links via primary and foreign keys to its process parameters (voltage, flow rate, tip-to-collector distance), material and solvent composition, ambient conditions (temperature, humidity), fiber outcomes (formation stability, diameter, diameter variation), structured morphology and instability annotations, image evidence, and the research metadata (DOI, title, abstract, device model where available). Mechanical and functional property fields (e.g., tensile strength, modulus, porosity, conductivity) are stored as optional descriptors. All numeric quantities are normalized through a shared unit reference table, giving a single canonical representation for each measured quantity.

Structurally variable descriptors use a hybrid relational-JSON design: multi-component compositions, process configurations, and needle/collector equipment pair a typed class field with a native JSON definition field, and image metadata (modality, magnification, scale) follows the same pattern. The JSON fields remain directly queryable, accommodating setup variability at low schema cost while preserving SQL filterability on class and image type. The complete schema, including table definitions, is archived as a versioned Zenodo release~\cite{mahdian_electrospinning-dataorg_2026-1}; publishing it as a version-controlled artifact provides an executable definition sufficient to reconstruct the database and lets the schema evolve under semantic versioning without invalidating prior releases.

\subsection{Application and Deployment Architecture}
The backend application layer is implemented using Spring Boot (Java) and mediates communication among the database, the React frontend, and the Python computational engine, exposing its functionality through RESTful APIs. It handles data submission, querying, and access control, and performs automated structural validation of submitted records.

The web-based user interface is developed in React.js and enables users to explore, query, and contribute data through a standardized interface. Users can filter experimental records based on material systems, process parameters, or morphological outcomes, and submit new records via structured forms aligned with the platform schema. All interactions between the frontend and backend are handled through JSON-based API requests, enabling consistent and reproducible access to the underlying data.
 
The application is deployed as a Docker-based containerized service and served through a reverse proxy (Nginx) over HTTPS, providing secure communication, efficient request routing, and production-grade stability. This layered architecture separates data storage, validation logic, and user interaction, ensuring scalability, maintainability, and reproducibility of the platform as the dataset evolves. Regular database backups are performed to ensure data persistence and prevent data loss.

\subsection{Data Contribution Pathways}
\label{sec:upload}
Contributors submit structured experimental records to Electrospinning-Data.org through three routes. The web submission form provides an authenticated, guided interface whose fields follow the platform's standardized schema, processing parameters, solution properties, environmental conditions, and morphology and fiber outcomes, and allows fiber images to be attached to a record; it requires no software installation on the contributor's machine. For batch entry, an Excel-based path parses spreadsheet files conforming to the platform's standardized field structure into structured records, so that multiple experiments can be ingested in a single submission. Programmatic contribution is available through a REST write API and an accompanying Python client, authenticated with personal access tokens. Regardless of route, submitted records then enter the curation lifecycle described in Section ~\ref{subsec:contribution} (Fig.~\ref{fig:lifecycle}).

\subsection{User Access, Authentication, and Roles}
Approved records are public, so browsing, searching, and downloading data require no account. Contributing or analyzing data requires authentication: contributors sign in to submit and manage their own records, and the Virtual Lab's analysis tools are available to authenticated users. Programmatic access is available through personal access tokens. Additional roles govern the platform itself, moderators review submitted records, and administrators manage broader operations, keeping contribution, curation, and governance clearly separated.

\subsection{Data Harmonization and Morphology Standardization}
\label{sec:harmonization}
Because Electrospinning-Data.org aggregates experiments reported across the published literature, the same quantities often appear in different units and fiber morphology is described in inconsistent free text. The platform reconciles this heterogeneity through the structured schema and canonical-unit normalization described in Section~\ref{sec:database}, together with a controlled morphology vocabulary that replaces free-text labels. The morphology vocabulary is published as a citable community reference on Zenodo~\cite{mahdian_cogni-emcv_2026} and is deliberately kept open for expansion, so that additional descriptors and categories can be incorporated as new fiber morphologies are reported. Because this harmonization is applied when records enter the platform rather than repeated for each analysis, data drawn from many independent studies can be searched, filtered, and analyzed together without per-source reconciliation.

\subsection{Data Validation and Expert Curation}
\label{subsec:contribution}
\begin{figure}[htbp]
    \centering
    \includegraphics[width=0.8\textwidth]{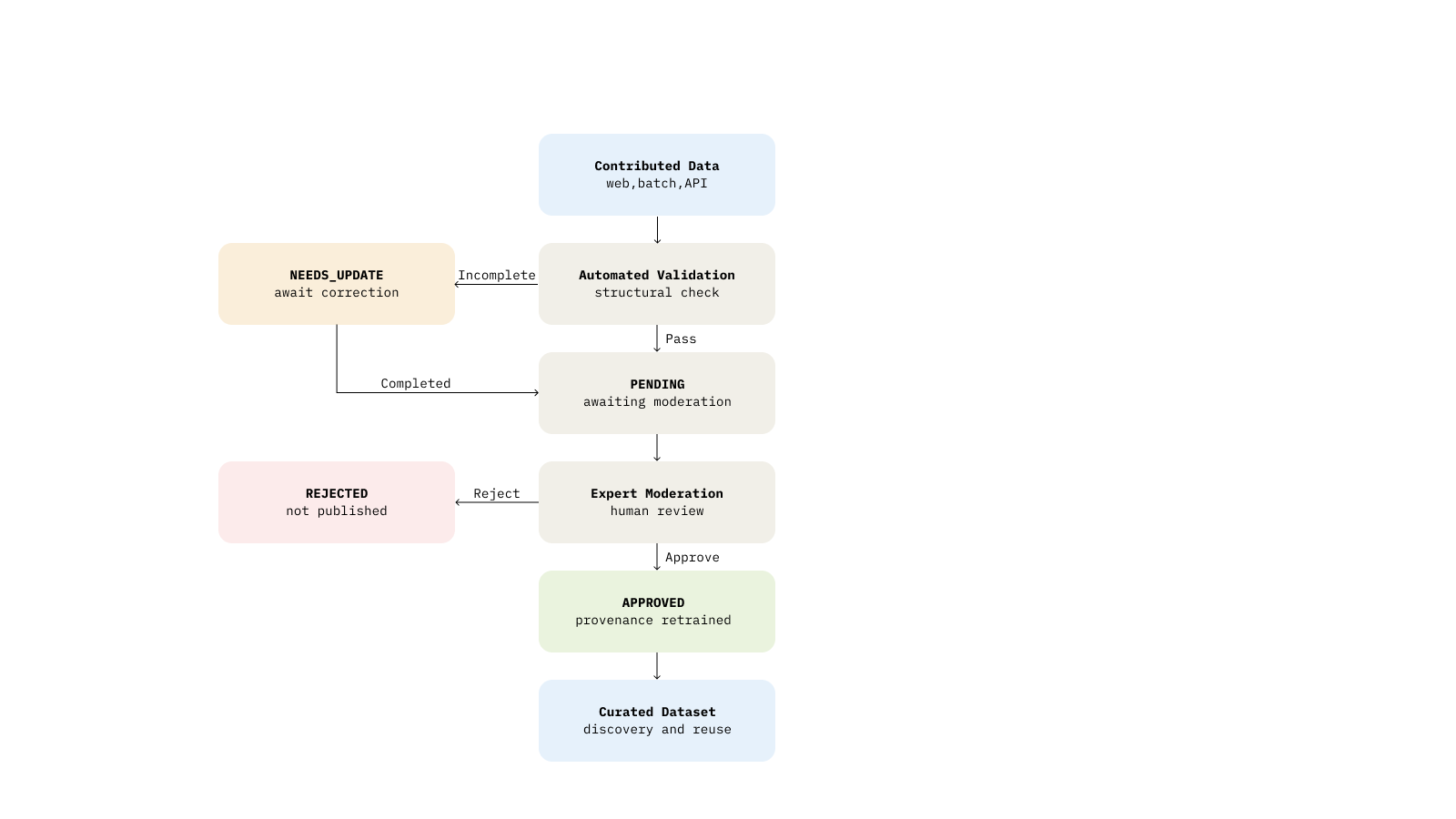}
    \caption{Data validation and curation lifecycle in Electrospinning-Data.org. Submitted records undergo automated structural validation: incomplete submissions are held as \texttt{NEEDS\_UPDATE} for correction, while those that pass are assigned \texttt{PENDING} for expert moderation. Moderators approve records into the public curated dataset, where they retain source and contributor provenance, or reject them from publication.}
    \label{fig:lifecycle}
\end{figure}

Submitted records enter a governed lifecycle with four states, \texttt{PENDING},\texttt{APPROVED}, \texttt{REJECTED}, and \texttt{NEEDS\_UPDATE}, before joining the public dataset (Fig.~\ref{fig:lifecycle}). Automated structural validation runs first, enforcing conformance to the record schema and field requirements; it governs a submission's structure and completeness, not the scientific correctness of its values. Records that pass are held as \texttt{PENDING} for review. Incomplete submissions are not discarded: a record missing required fields is preserved as \texttt{NEEDS\_UPDATE}, with the missing fields reported back so the contributor can complete it and return it to \texttt{PENDING}, a progressive-completion path suited to programmatic and machine-generated workflows, whereas browser submissions apply stricter validation at entry. Pending records are then reviewed by human moderators. Approved records become public and retain their source and contributor provenance; rejected records are not published. Both the automated check and the moderators' review draw on a published set of validation and verification rules~\cite{mahdian_cogni-evvr_2026}. Together, automated validation and expert moderation govern data structure and admission to the curated dataset, while scientific correctness continues to depend on the underlying sources and curator judgment.

\subsection{Versioned Dataset Releases and Archival Deposition}

Electrospinning-Data.org issues periodic versioned releases of the curated dataset. Once newly approved contributions have accumulated, a release is prepared as a stable, identifiable snapshot with version metadata such as record count, creation time, and license, retrievable from the platform in spreadsheet and archive formats. Each snapshot is deposited to Zenodo, which assigns a persistent digital object identifier (DOI), and is distributed under the Creative Commons Attribution 4.0 International (CC~BY~4.0) license. The DOI is recorded in the version metadata and exposed within the platform, so an analysis can be tied to a specific, citable snapshot. This separation between the operational live dataset and discrete, DOI-bearing snapshots supports reproducible referencing and long-term citation of the data underlying downstream work.

\subsection{Integrated Analytical Environment: The Virtual Lab}
The Virtual Lab is an authenticated computational workspace that couples the curated dataset to the analysis and modeling services of the Python engine, so that researchers can analyze and model the curated data within the platform itself. Its modules share a single working-dataset context: a researcher first assembles a working dataset through an interactive query builder with filtering and missing-data handling, in which identifiers and source metadata are held separate from analytical variables so that record identifiers are not treated as measurements, and this dataset then carries through every subsequent module, exploratory data analysis, supervised machine-learning training and evaluation, model retention and reuse, model interpretation, and evidence-based parameter exploration. Each stage records the inputs and settings used to produce its outputs, keeping computational results traceable to the specific working dataset they were derived from. The individual modules and representative analyses are presented in Section~\ref{sec:results}.

\clearpage
\section{Results}
\label{sec:results}
Electrospinning-Data.org integrates data sharing and computational analysis within a single research infrastructure. Its main services comprise a public data repository for the contribution, discovery, and retrieval of structured electrospinning records, together with the Virtual Lab for constructing working datasets, exploratory data analysis, machine-learning modeling and interpretation, and evidence-based parameter exploration. Table~\ref{tab:platform_overview} provides an overview of these services, their intended users, objectives, and access requirements.

\begin{table*}[htbp]
\centering
\caption{Overview of the main services of Electrospinning-Data.org.}
\label{tab:platform_overview}
\renewcommand{\arraystretch}{1.25}
\begin{tabularx}{\textwidth}{p{3.0cm} X X p{3.0cm}}
\hline
\textbf{Section} & \textbf{Content} & \textbf{Objective} & \textbf{Access} \\
\hline

\textbf{Data Repository} &
Experimental records; upload, search, download and API access &
FAIR collection and dissemination of electrospinning data &
Open; registration for contribution \\

\textbf{Dataset Builder} &
Data filtering, selection and export &
Construction of analysis-ready datasets &
Registration required \\

\textbf{Exploratory Data Analysis} &
Statistics, correlations, dimensionality reduction and outlier analysis &
Interactive data exploration &
Registration required \\

\textbf{ML \& Explainability} &
Model training, evaluation, reuse, inference and SHAP analysis &
Predictive modelling and interpretation &
Registration required \\

\textbf{Recipe Builder} &
Similar-record retrieval and evidence-linked parameter configurations &
Evidence-based parameter exploration &
Registration required \\

\hline
\end{tabularx}
\end{table*}

\subsection{Public Data Access and Retrieval}

The public electrospinning dataset can be accessed through three
complementary read channels. All operate over the same underlying data
resource and expose the platform's version model, so researchers can
retrieve either the continuously updated live dataset or a specific,
immutable dataset version. Requesting a named version rather than the
moving live dataset ties downstream analyses to a defined release,
providing a reproducibility anchor at the data level.

\begin{figure*}[t]
    \centering

    \includegraphics[width=\textwidth]
    {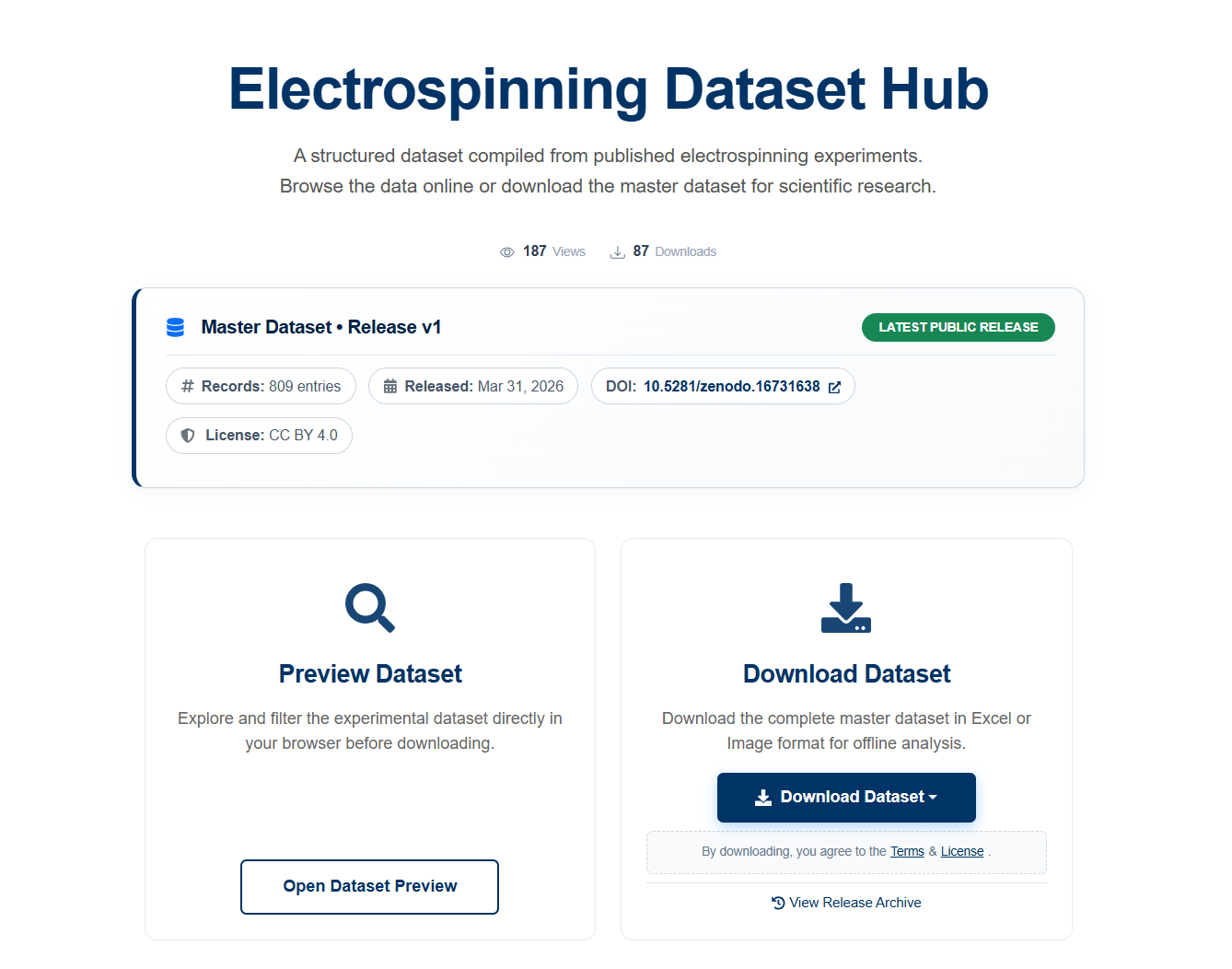}

    \textbf{(a) Web interface}

    \vspace{0.8em}

    \includegraphics[width=\textwidth]
    {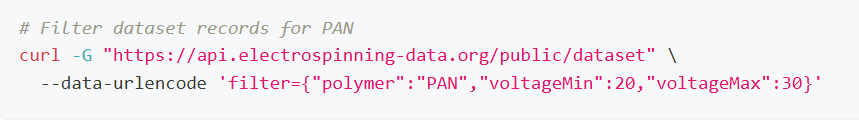}

    \textbf{(b) REST API}

    \vspace{0.8em}

    \includegraphics[width=\textwidth]
    {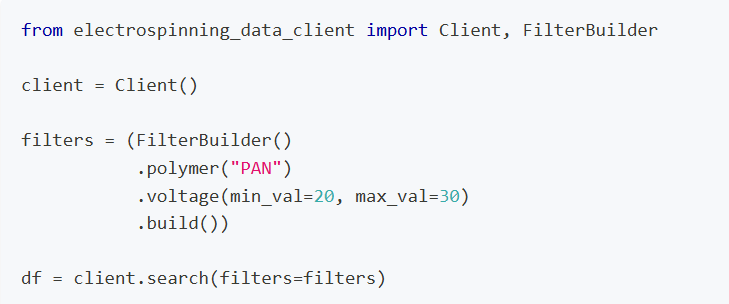}

    \textbf{(c) Python client}

    \caption{\textbf{Complementary access routes to the
    Electrospinning-Data.org dataset.}
    (a) The web-based Dataset Hub provides interactive dataset preview and
    download together with release metadata, including the dataset version,
    persistent DOI, license, and release archive.
    (b) The public REST API enables programmatic filtering and retrieval of
    experimental records; the example selects PAN records with an applied
    voltage between 20 and 30~kV.
    (c) The Python client provides a higher-level programmatic interface
    implementing the equivalent query for integration into Python-based
    scientific workflows. All three interfaces operate over the same
    underlying structured data resource.}
    \label{fig:data_access}

\end{figure*}

\begin{enumerate}

    \item \textbf{Web interface.}
    The dataset can be explored and downloaded through the
    Electrospinning-Data.org Dataset Hub without programming tools.
    Researchers can preview experimental records, filter them into subsets,
    and download the result. The Hub also exposes release metadata-dataset
    version, record count, release date, persistent DOI, license, and access
    to previous releases (Fig.~\ref{fig:data_access}a).

    \item \textbf{REST API.}
    The public REST API provides programmatic access with support for
    filtering, pagination, and dataset-version selection, enabling external
    software and scripted workflows to retrieve either the complete dataset
    or subsets satisfying specified experimental criteria. An example query
    filtering polyacrylonitrile (PAN) records by applied voltage is shown in
    Fig.~\ref{fig:data_access}b.

    \item \textbf{Python client library.}
    The \texttt{electrospinning-data-client} package provides a higher-level
    interface to the REST API for integration into Python-based scientific
    workflows, supporting retrieval, filtering, and dataset-version selection
    without constructing HTTP requests directly. The equivalent PAN query is
    shown in Fig.~\ref{fig:data_access}c.

\end{enumerate}

Across all three channels, data are available in machine-readable formats
including JSON and Excel (XLSX), and associated fiber images can be
retrieved as ZIP archives where available.

\subsection{Implementation of the FAIR Principles}

The FAIR principles are operationalized across the platform through
complementary mechanisms for data discovery, access, standardization,
provenance, and preservation. Table~\ref{tab:fair_implementation}
summarizes the principal platform features supporting each FAIR dimension.

\begin{table}[htbp]
\centering
\caption{Implementation of the FAIR principles in Electrospinning-Data.org.}
\label{tab:fair_implementation}
\begin{tabular}{p{0.16\textwidth} p{0.76\textwidth}}
\hline
\textbf{FAIR principle} & \textbf{Implementation in Electrospinning-Data.org} \\
\hline

\textbf{Findable} &
Experimental records are represented using structured, searchable metadata
and can be discovered through the web interface and programmatic queries.
Periodic dataset releases are deposited in Zenodo and assigned persistent
DOIs. \\[0.5em]

\textbf{Accessible} &
Approved records are publicly accessible without authentication through the
web platform, REST API, and Python client. Data can be retrieved as complete
datasets or filtered subsets, including specific versioned releases, in
machine-readable downloadable formats. \\[0.5em]

\textbf{Interoperable} &
Experimental observations follow a common domain-specific schema with
standardized field definitions, canonical units, structured JSON
representations for variable experimental configurations, and controlled
morphology descriptors. REST APIs and the Python client provide
machine-readable interfaces for integration with external computational
workflows. \\[0.5em]

\textbf{Reusable} &
Records retain source and contributor provenance together with experimental
and research metadata. Curated datasets are distributed under an explicit
license (CC BY 4.0), while immutable DOI-bearing releases preserve defined dataset states
that can be cited and reused in reproducible downstream analyses. \\

\hline
\end{tabular}
\end{table}

\clearpage

\subsection{Illustrative Analyses in the Virtual Lab}

The Virtual Lab allows a researcher to carry an electrospinning dataset
from selection through exploratory analysis, supervised modelling and
model interpretation without exporting data to a local environment.
Because the working-dataset context is shared across modules, a subset
defined during dataset selection propagates to the exploratory,
modelling and explainability stages, and each stage records the inputs
and settings used to produce its outputs. The examples below illustrate
this workflow on the live dataset; they are intended to demonstrate the
analytical environment rather than to establish validated models or
scientific findings.

\subsubsection{Exploratory Data Analysis}

The exploratory data analysis (EDA) module operates on the selected
working dataset and provides several coordinated views, distribution,
relationship, correlation, dimensionality reduction, feature quality,
outlier detection and statistical summary, that can be inspected before
any modeling. In the example shown here the working dataset is the live
Master Dataset (809 records).

\begin{figure}[htbp]
  \centering
  \includegraphics[width=\linewidth]{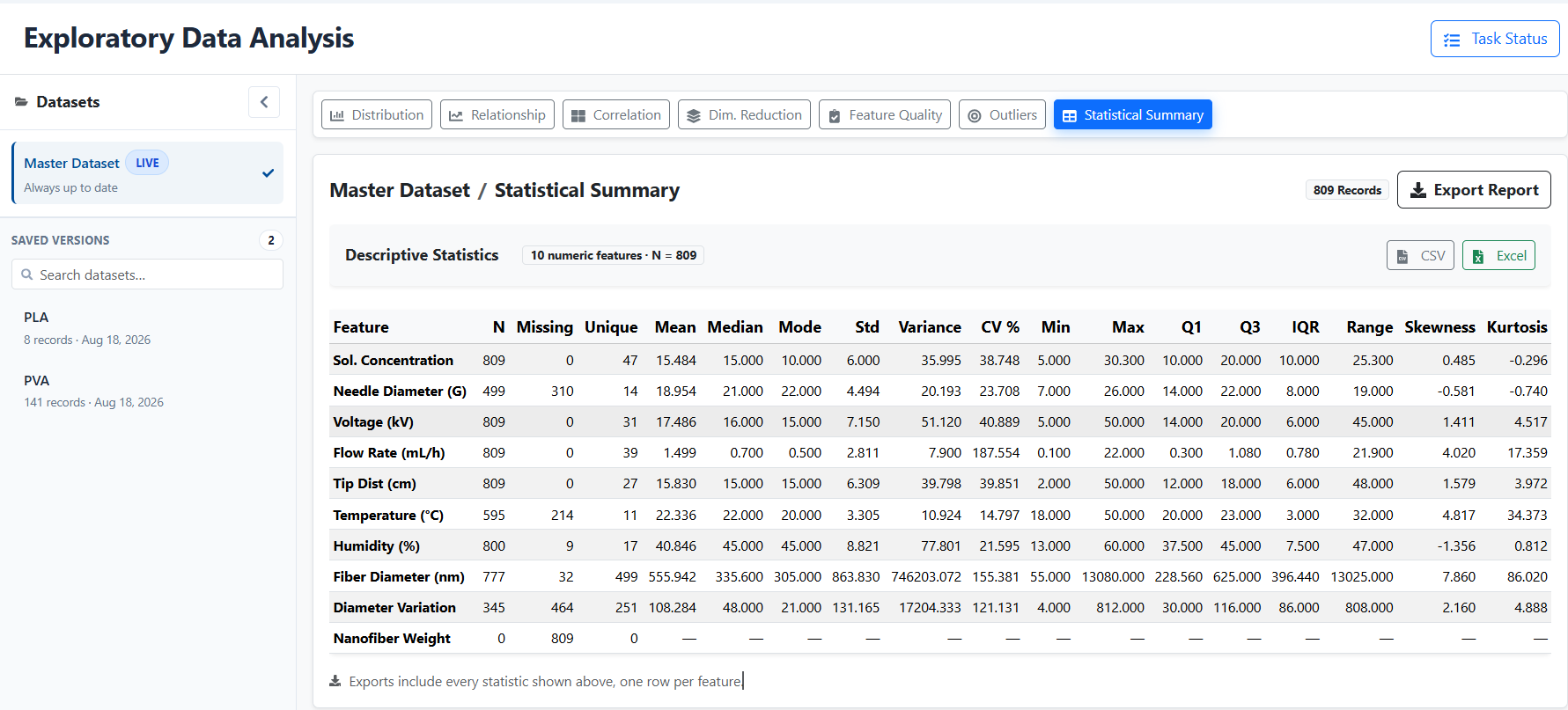}
  \caption{Statistical-summary view of the exploratory data analysis
  module, computed on the live Master Dataset (809 records). For each
  numeric feature the module reports non-missing count, missing and
  unique counts, central tendency (mean, median, mode), dispersion
  (standard deviation, variance, coefficient of variation, interquartile
  range, range and extrema) and distribution shape (skewness, kurtosis).}
  \label{fig:vlab_eda_stats}
\end{figure}

\paragraph{Descriptive Statistics.}
The statistical summary reports per-feature descriptive statistics for
the numeric variables in the working dataset, including non-missing
count, missing and unique counts, measures of central tendency (mean,
median, mode), dispersion (standard deviation, variance, coefficient of
variation, interquartile range, range and extrema) and distribution
shape (skewness and kurtosis) (Fig.~\ref{fig:vlab_eda_stats}).
Presenting these statistics together makes two properties of the live
dataset immediately visible. First, feature coverage is uneven: the core
process and solution parameters (solution concentration, applied
voltage, flow rate and tip-to-collector distance) are fully populated
across the 809 records, whereas environmental variables and the
fiber-diameter variation outcome are less complete. Second, the fiber-diameter
outcome is strongly right-skewed and heavy-tailed, with its mean well
above its median, consistent with the order-of-magnitude spread of
diameters observed in the reference dataset~\cite{mahdian_cogni-e-spindb_2026}. A report in different format can be generated and the
summary can be exported as CSV or Excel for external reporting.

\begin{figure}[htbp]
  \centering
  \includegraphics[width=\linewidth]{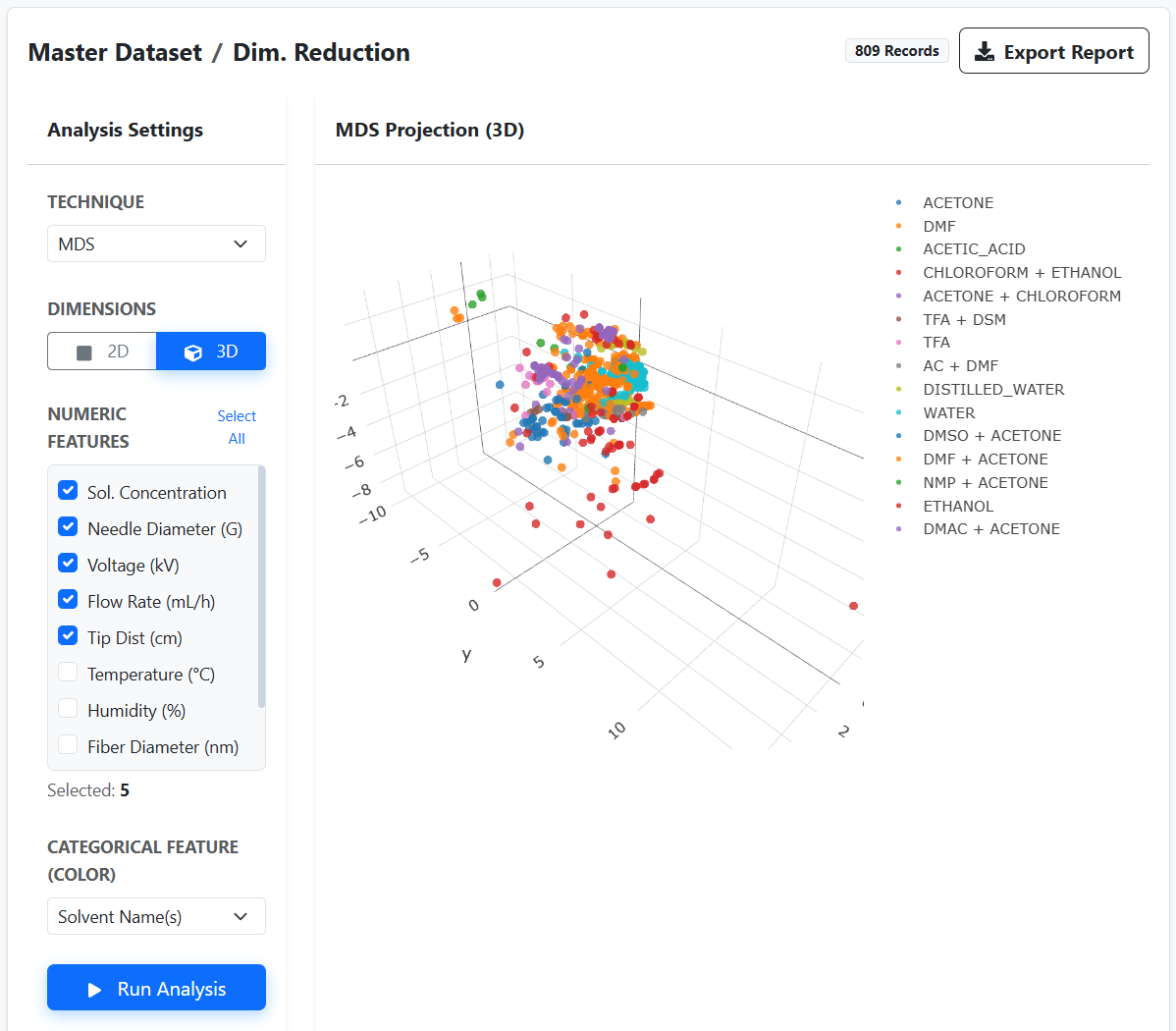}
  \caption{Three-dimensional multidimensional-scaling (MDS) projection
  produced by the dimensionality-reduction view of the EDA module,
  computed on the live Master Dataset (809 records) from a selected
  subset of process and solution parameters. Each point represents an
  experimental record; points are colored by solvent system. The view
  also offers principal component analysis and kernel PCA, t-SNE, UMAP
  and Isomap, with two- or three-dimensional output.}
  \label{fig:vlab_eda_dimred}
\end{figure}

\paragraph{Dimensionality Reduction.}
The dimensionality-reduction view projects the multivariate numeric
feature space onto two or three dimensions for visual inspection of
structure that is not apparent in univariate summaries. The module
supports several linear and non-linear techniques, principal component
analysis and kernel PCA, t-SNE, UMAP, multidimensional scaling (MDS) and
Isomap, and the researcher selects the numeric features to include and
an optional categorical variable used to color the projected points.
Figure~\ref{fig:vlab_eda_dimred} shows a three-dimensional MDS projection
of the live Master Dataset computed from a selected subset of process and
solution parameters, with points colored by solvent system. The
projection makes the chemical diversity of the corpus visible, single-
and binary-solvent systems spanning a range of chemistries, and lets a
researcher inspect whether records group by solvent before modeling.

\begin{figure}[htbp]
  \centering
  \includegraphics[width=\linewidth]{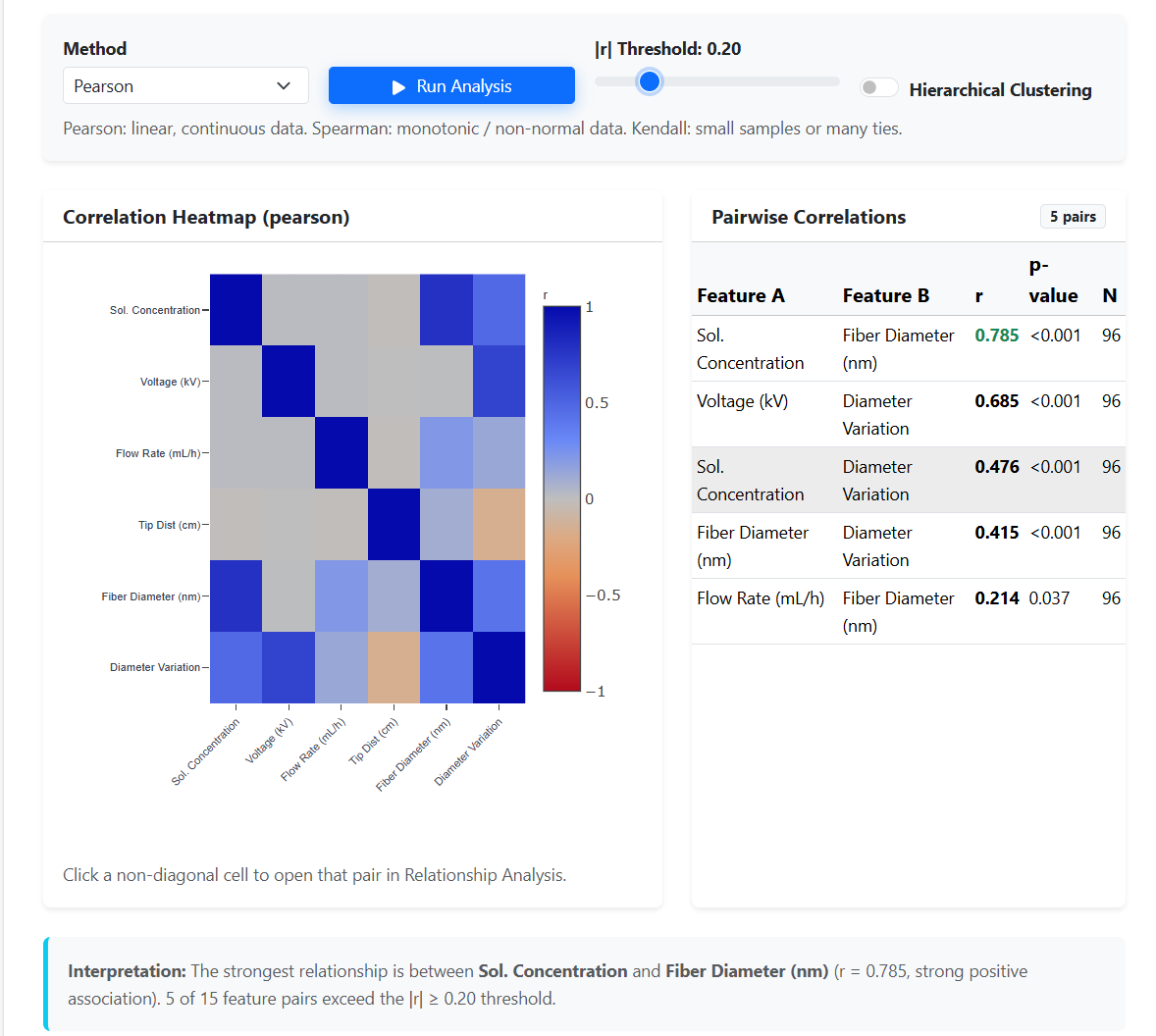}
  \caption{Correlation view of the EDA module, computed on a saved
  dataset version (the poly(vinyl alcohol) (PVA) subset, 96 records). A
  Pearson correlation heatmap over six numeric features (left) is paired
  with a ranked table of feature pairs exceeding an absolute-correlation
  threshold of $0.20$ (right), each reported with correlation
  coefficient, p-value and sample size. The module also offers Spearman
  and Kendall methods and optional hierarchical clustering; selecting a
  non-diagonal heatmap cell opens the corresponding pair in the
  relationship view.}
  \label{fig:vlab_eda_corr}
\end{figure}

\begin{figure}[htbp]
  \centering
  \includegraphics[width=\linewidth]{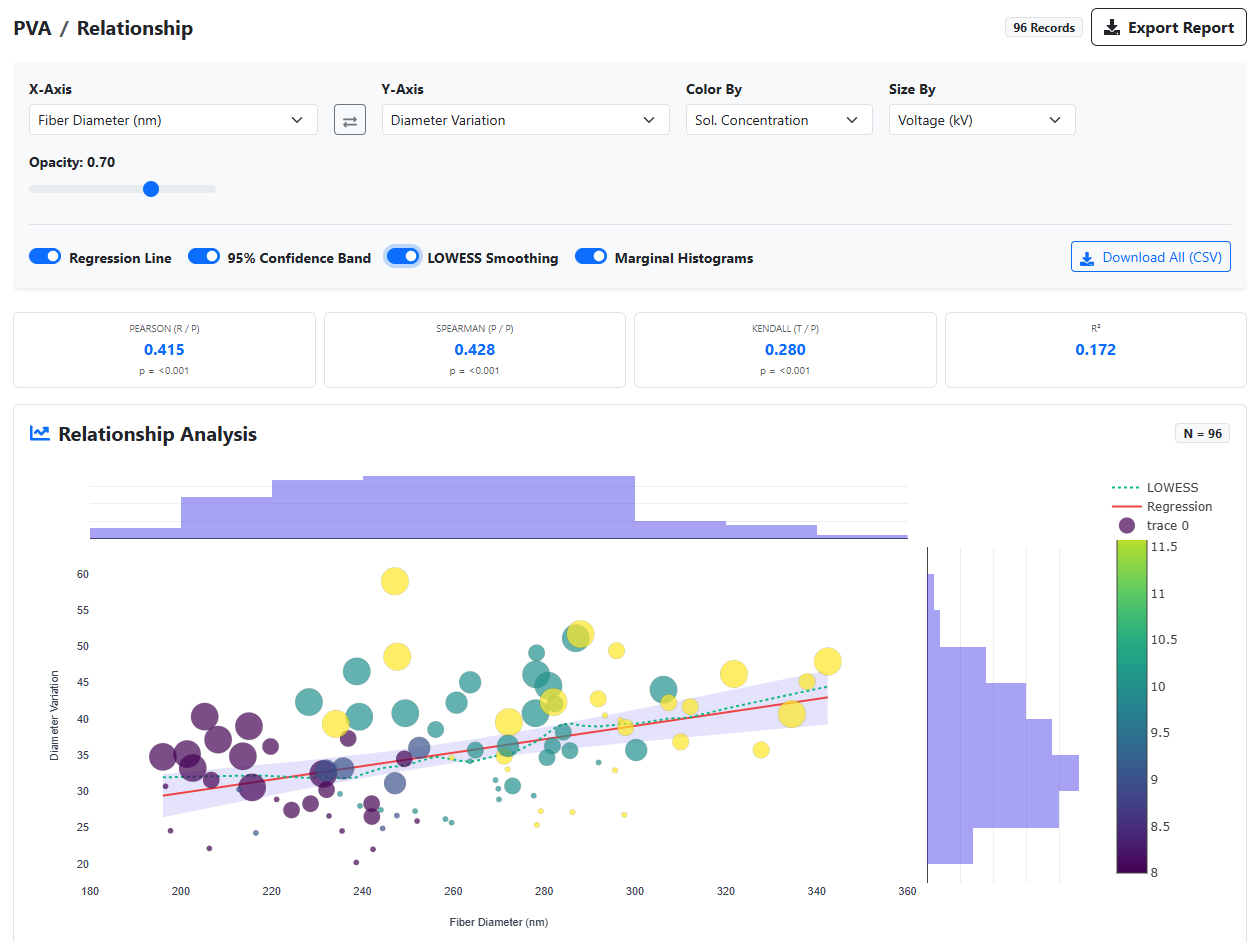}
  \caption{Pairwise relationship view, opened by selecting a non-diagonal
  cell of the correlation heatmap. Fiber diameter is plotted against
  diameter variation for the same saved PVA subset (96 records), with
  solution concentration mapped to point colour and applied voltage to
  point size. A regression line with 95\% confidence band, LOWESS
  smoothing and marginal histograms on both axes are enabled, and axis
  assignment and point opacity are adjustable. Pearson, Spearman and
  Kendall coefficients with their p-values and the coefficient of
  determination ($R^{2}$) are reported above the plot, and the underlying
  records can be exported as CSV. The reported coefficients are
  descriptive measures of association and are not, in themselves, evidence
  of causal relationships.}
  \label{fig:vlab_eda_rel}
\end{figure}

\paragraph{Correlation Analysis.}
The correlation view computes pairwise correlations among the numeric
features of the working dataset. The researcher chooses the correlation
method, Pearson, Spearman or Kendall, with in-app guidance on when each
is appropriate, sets an absolute-correlation threshold, and can
optionally apply hierarchical clustering to reorder the correlation
matrix. The result combines a correlation heatmap with a ranked table of
feature pairs that exceed the threshold, each reported with its
correlation coefficient, significance (p-value) and sample size, together
with a short interpretation of the strongest association
(Fig.~\ref{fig:vlab_eda_corr}). Selecting a non-diagonal heatmap cell
opens that pair in the relationship view, linking the correlation
overview to closer pairwise inspection (Fig.~\ref{fig:vlab_eda_rel}). In
the example shown, a Pearson analysis is applied not to the live dataset
but to a saved dataset version corresponding to the poly(vinyl alcohol)
(PVA) subset (96 records), over six numeric variables; it identifies
solution concentration and fiber diameter as the most strongly associated
pair. As with the other exploratory views, these coefficients are
descriptive measures of association on the selected working dataset; the
reported p-values and sample sizes make the statistical basis explicit
and are not, in themselves, evidence of causal relationships.

\clearpage

\begin{figure}[htbp]
  \centering
  \includegraphics[width=\linewidth]{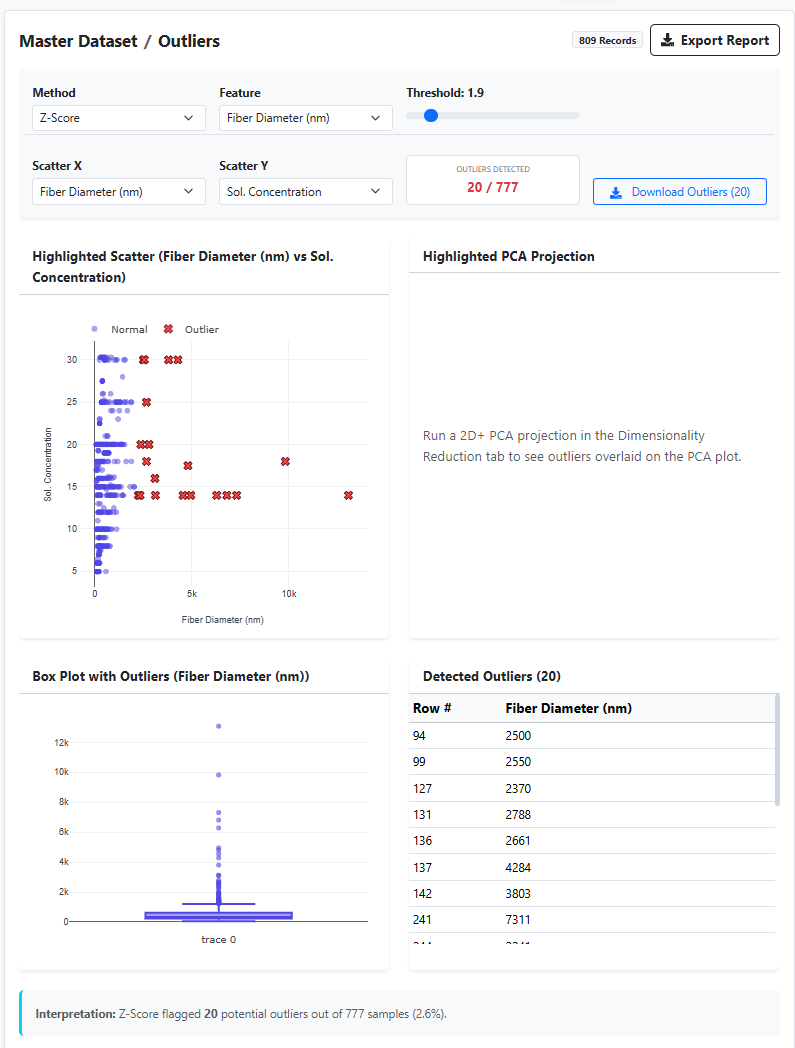}
  \caption{Outlier view of the EDA module, applied to fiber diameter on
  the live Master Dataset. A Z-score criterion (threshold $1.9$) flags
  $20$ of $777$ records ($\approx 2.6\%$) as potential outliers, shown as
  a highlighted scatter of fiber diameter against solution concentration
  (top left), a box plot of fiber diameter (bottom left) and a table of
  the flagged records (bottom right). Flagged records are exported for inspection rather than removed.}
  \label{fig:vlab_eda_outliers}
\end{figure}

\paragraph{Outlier and Anomaly Detection.}
The outlier view flags candidate anomalous records in the working
dataset. The researcher selects a detection method, a statistical
criterion such as a Z-score, together with the
feature to examine and a sensitivity threshold. The results are presented
as a highlighted scatter plot distinguishing flagged from unflagged
records, a box plot of the examined feature, and a table of flagged
records with their values; the flagged set can be downloaded, and can
additionally be overlaid on the PCA projection produced in the
dimensionality-reduction view. In the example shown, a Z-score criterion
applied to fiber diameter on the live Master Dataset flags 20 of 777
records (approximately $2.6\%$) as potential outliers, concentrated at
the large-diameter end of the distribution
(Fig.~\ref{fig:vlab_eda_outliers}). Consistent with the module's design,
such records are flagged for inspection and export rather than removed
automatically; extreme fiber diameters need not indicate erroneous data
and may reflect genuine variability, in line with the treatment of such
values as tail observations rather than invalid measurements in the
reference dataset~\cite{mahdian_cogni-e-spindb_2026}.

\subsubsection*{Machine-learning training}

The machine-learning module lets a researcher train and evaluate
supervised models on the working dataset through a guided, multi-step
workflow: selecting the objective and target, choosing features,
specifying missing-value handling, scaling and categorical encoding,
setting the model and its parameters, reviewing the configuration, and
inspecting results. Because each step is explicit, the preprocessing
decisions are retained together with the trained model rather than left
implicit. The module supports both regression and classification and
offers a range of algorithms, linear and regularized regression, tree
ensembles and gradient-boosting methods (including Random Forest,
XGBoost, CatBoost and Gradient Boosting), and support-vector and
nearest-neighbour models, among others, together with several
evaluation workflows: a train/test split, $k$-fold cross-validation,
whole-data training, and hyperparameter tuning by grid or randomized
search. Regression models are scored with $R^2$, MAE and RMSE, and
classification models with accuracy, F1, precision, recall and ROC-AUC.

\begin{figure}[htbp]
  \centering
  \includegraphics[width=\linewidth]{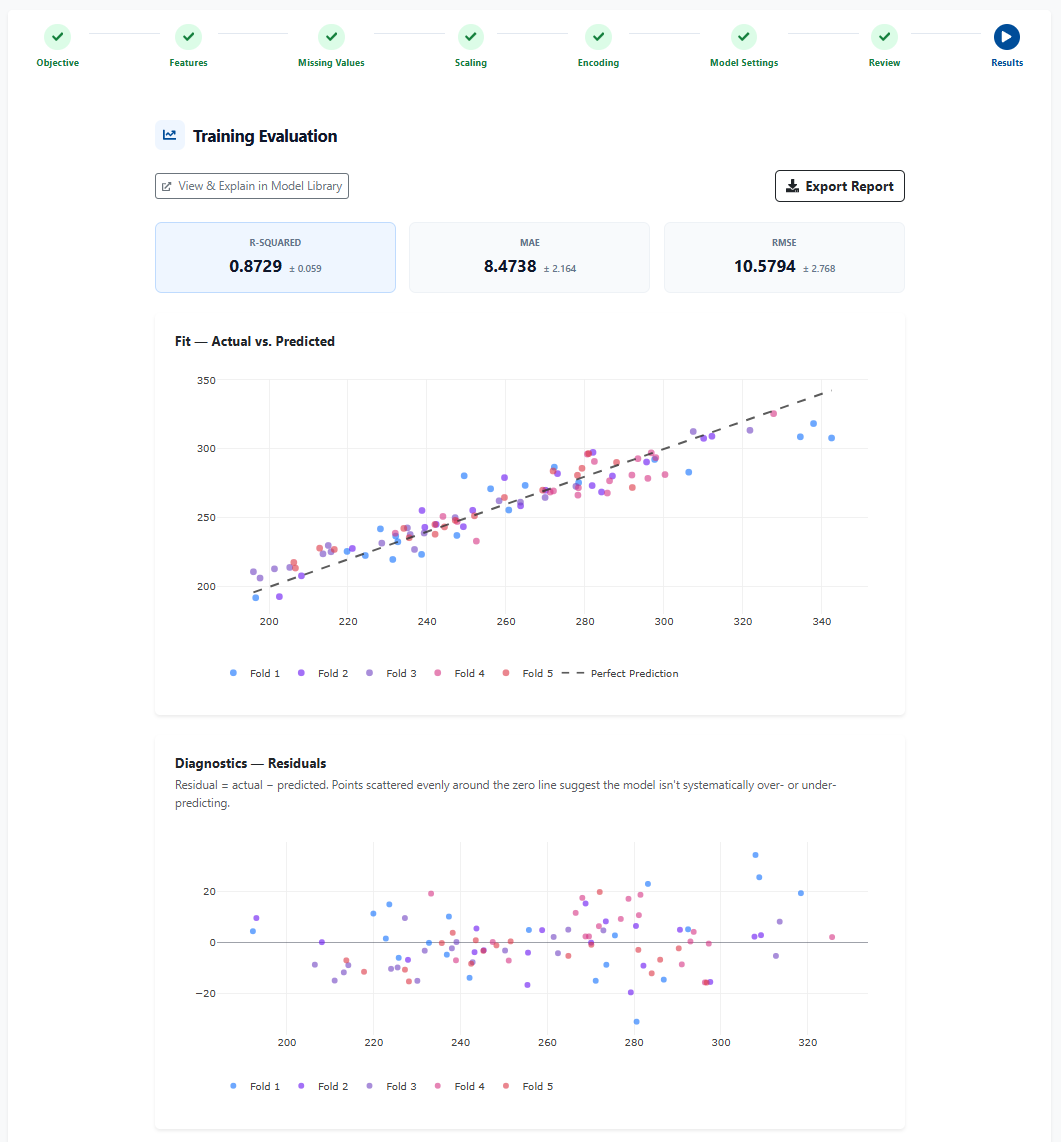}
  \caption{Training-evaluation result from the machine-learning module
  for a gradient-boosting regressor (100 estimators, learning rate $0.1$,
  maximum depth $3$) predicting fiber diameter from applied voltage, flow
  rate, tip-to-collector distance and solution concentration on the saved
  PVA dataset version, evaluated by five-fold cross-validation. The panel
  reports cross-validated $R^2 = 0.873 \pm 0.059$,
  MAE $= 8.47 \pm 2.16$~nm and RMSE $= 10.58 \pm 2.77$~nm
  (mean $\pm$ standard deviation across folds), an actual-versus-predicted
  fit against the perfect-prediction line (points colored by fold) and a
  residual diagnostic plot.}
  \label{fig:vlab_ml_training}
\end{figure}

Figure~\ref{fig:vlab_ml_training} shows one such run on the saved PVA
dataset version used above (Fig.~\ref{fig:vlab_eda_corr} and Fig.~\ref{fig:vlab_eda_rel}). A
gradient-boosting regressor (100 estimators, learning rate $0.1$,
maximum depth $3$) was trained to predict fiber diameter from four
process and solution parameters-applied voltage, flow rate,
tip-to-collector distance and solution concentration, and evaluated by
five-fold cross-validation. The evaluation panel reports the
cross-validated $R^2$, MAE and RMSE with their across-fold variability,
an actual-versus-predicted fit against the perfect-prediction line
colored by fold, and a residual diagnostic plot. A ``View \& Explain in
Model Library'' action carries the trained model into the model library
for reuse and explanation. These metrics summarize the performance of
this particular model on the selected working dataset under
cross-validation; they demonstrate the training and evaluation
functionality and are not, in themselves, evidence of predictive accuracy
or of generalization to other materials, laboratories or datasets.

\subsubsection{Model library}

\begin{figure}[htbp]
  \centering
  \includegraphics[width=\linewidth]{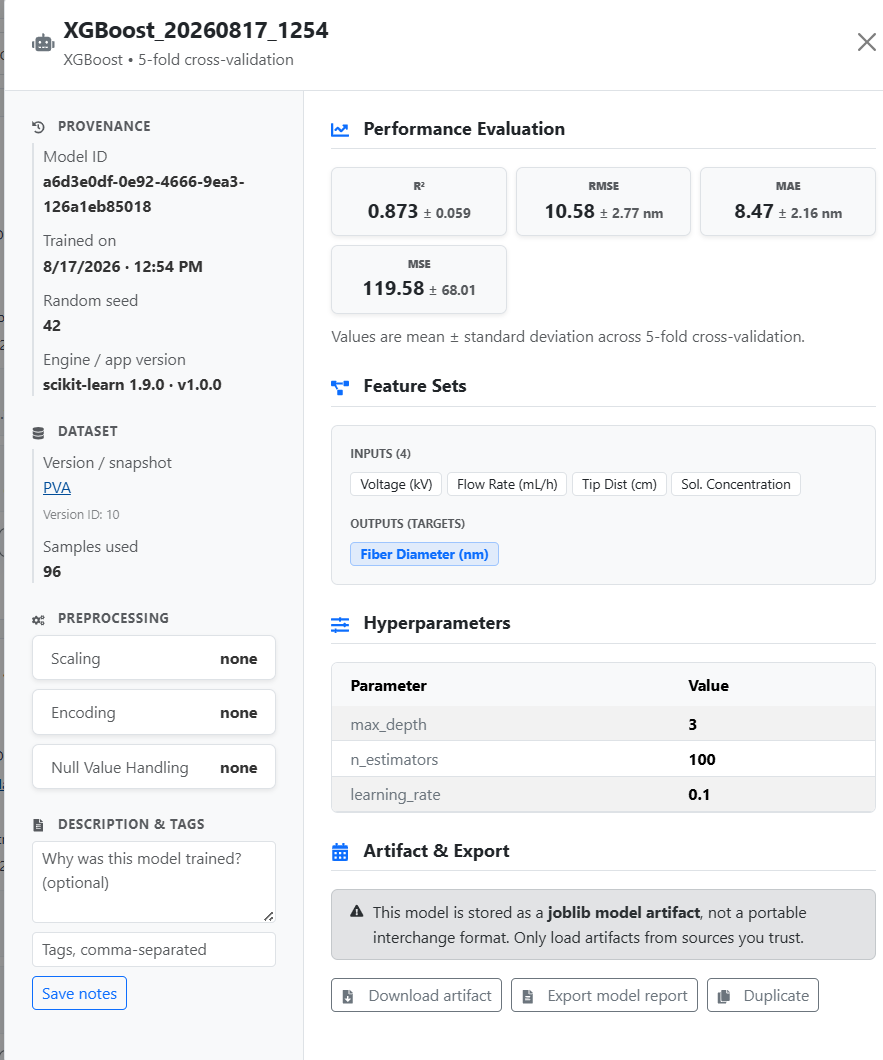}
  \caption{Model-library detail view for a stored model (an XGBoost
  regressor trained on the PVA dataset snapshot). The provenance panel
  (left) records the model identifier, training timestamp, random seed
  ($42$), engine and application version (scikit-learn 1.9.0), the dataset
  version or snapshot and sample count (PVA, version ID 10, 96 samples),
  and the preprocessing choices. The evaluation panel (right) reports the
  five-fold cross-validation metrics, the four input features and the
  target, and the model hyperparameters, and provides artefact download,
  model-report export and duplication actions. The interface notes that
  the stored artefact is a \texttt{joblib} file rather than a portable
  interchange format and should be loaded only from trusted sources.}
  \label{fig:vlab_model_library}
\end{figure}

Trained models are retained in a model library, which presents them as a
searchable and comparable collection: models can be located by name,
algorithm, target or dataset, filtered and sorted, and placed side by
side for comparison. Each entry records the metadata needed to identify
and reproduce the run, a unique model identifier, the training
timestamp, the random seed, the engine and application version, the
dataset version or snapshot together with its version identifier and the
number of samples used, the preprocessing choices, the input features
and target, and the model hyperparameters
(Fig.~\ref{fig:vlab_model_library}). From the library a stored model can
be reused without retraining: it can be evaluated on a single
user-specified input to return a prediction, duplicated as the basis for
a new configuration, passed to the explainability module, or exported.

Models can be exported as a serialised \texttt{joblib} artefact. The library therefore functions as a working record of models trained on
the platform, spanning different algorithms, dataset versions and
validation schemes, rather than a curated set of endorsed results, and
retention of a model is not in itself a performance claim.

\subsubsection*{Model explainability}

\begin{figure}[htbp]
    \centering

    \begin{subfigure}[t]{\linewidth}
        \centering
        \includegraphics[width=\linewidth]{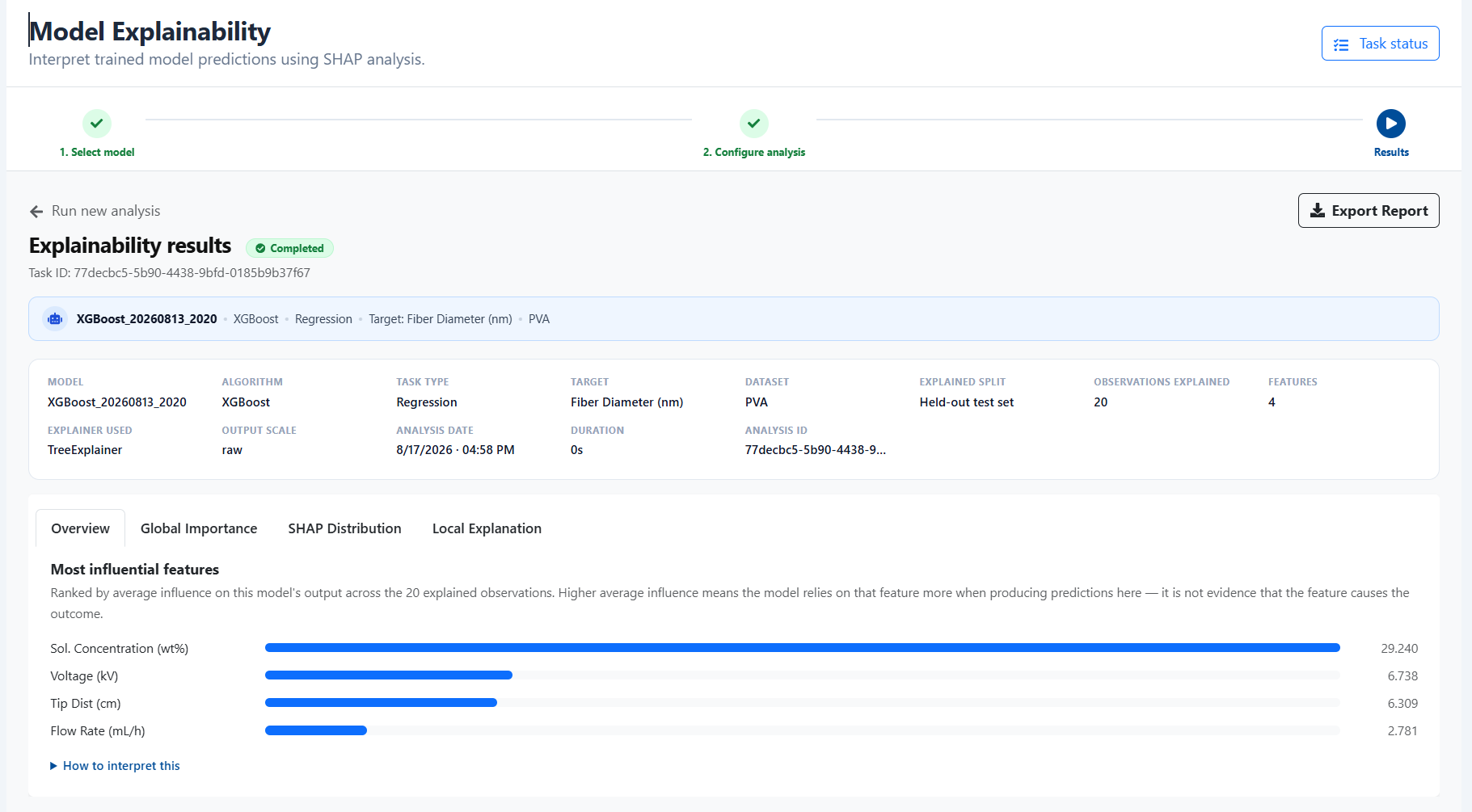}
        \caption{}
        \label{fig:vlab_explainability_a}
    \end{subfigure}

    \vspace{1ex}

    \begin{subfigure}[t]{\linewidth}
        \centering
        \includegraphics[width=\linewidth]{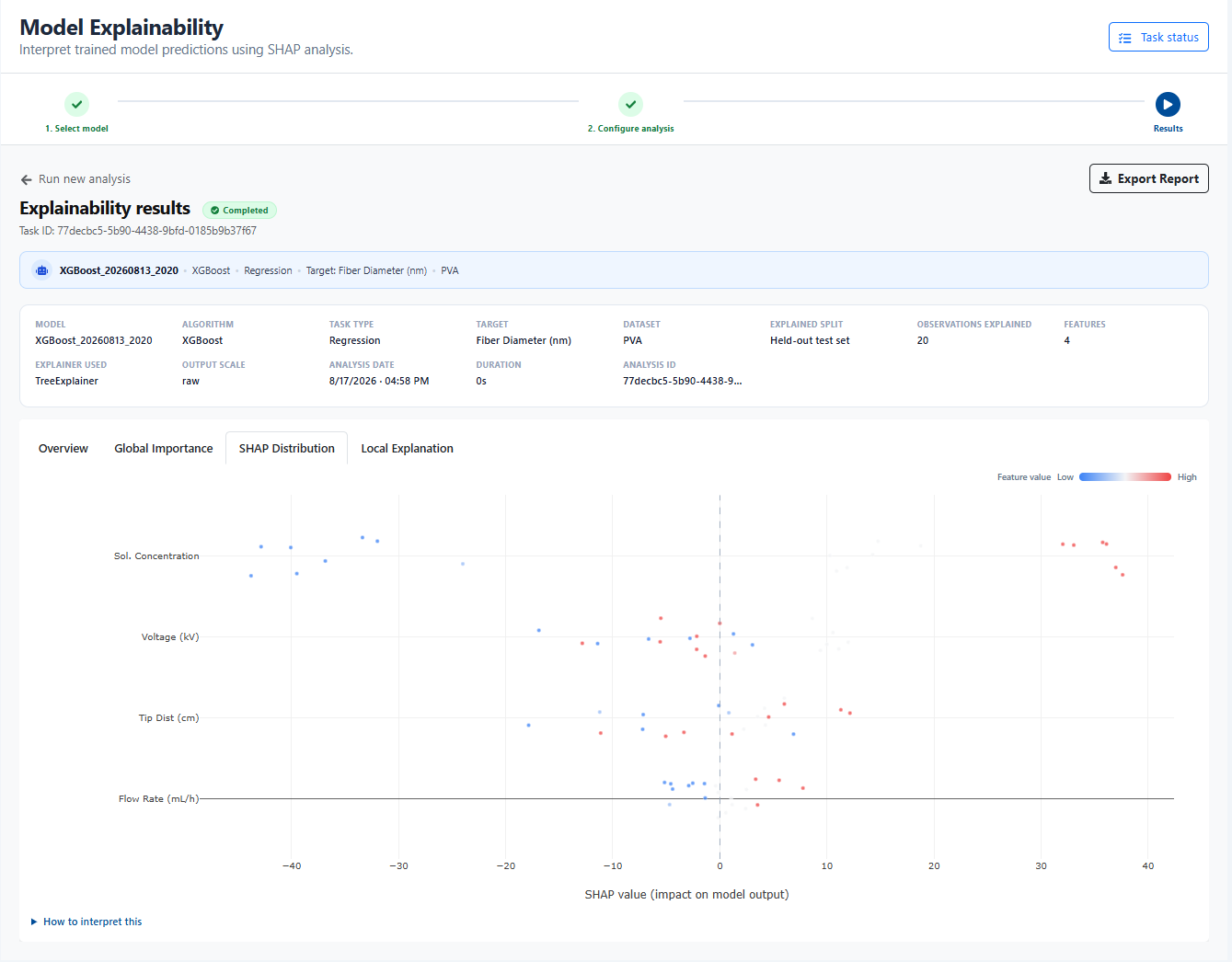}
        \caption{}
        \label{fig:vlab_explainability_b}
    \end{subfigure}

    \caption{Model-explainability result from the SHAP-based interpretation
    module, for the XGBoost model trained on the PVA subset to predict fiber
    diameter, computed with a TreeExplainer on the 20 held-out test
    observations of a train/test-split model with four input features.
    (a)~Overview ranking features by their average absolute SHAP value
    (average influence on the model's output): solution concentration
    ($29.24$) dominates, followed by applied voltage ($6.74$),
    tip-to-collector distance ($6.31$), and flow rate ($2.78$).
    (b)~SHAP-distribution (beeswarm) plot: each point is one held-out
    observation, its horizontal position giving that feature's signed
    contribution to the prediction and its color the observation's feature
    value (low to high). As the interface notes, these views describe how
    strongly, and in which direction, the model relies on each feature for
    its predictions on the explained observations, and are not evidence of
    physical causation.}
    \label{fig:vlab_explainability}
\end{figure}

The explainability module interprets the predictions of a trained model
using SHAP (SHapley Additive exPlanations). Because SHAP values are
computed on a held-out test set, the module operates on models trained
with a train/test split; it reports the explainer used (a TreeExplainer
for the tree-based model shown), the output scale, the number of
observations explained and the features considered, together with an
analysis identifier and timestamp
(Fig.~\ref{fig:vlab_explainability}). The results are organised into
several views: an overview that ranks the most influential features by
their average absolute SHAP value (Fig.~\ref{fig:vlab_explainability}a), a
global-importance summary, a SHAP-distribution (beeswarm) plot showing
the per-observation contribution of each feature colored by that
observation's feature value (Fig.~\ref{fig:vlab_explainability}b), and a
local-explanation view for individual predictions.

In the example shown, an XGBoost model trained on the PVA subset to
predict fiber diameter from the same four process and solution parameters
is explained on its 20 held-out test observations. Solution concentration
is, by a wide margin, the feature the model relies on most, followed by
applied voltage, tip-to-collector distance and flow rate, an ordering
consistent with the strong concentration fiber-diameter association
identified in the correlation view (Fig.~\ref{fig:vlab_eda_corr}), while
the beeswarm plot further resolves the direction of these contributions
across individual observations. The interface states explicitly, and we
reiterate, that these values quantify how much the model relies on each
feature when producing predictions on the explained observations; they
describe the model's behavior on this working dataset and are not
evidence that any feature physically causes the fiber-diameter outcome.

\subsubsection{Evidence-Based Parameter Exploration: Recipe Builder}

The Recipe Builder assembles a data-informed starting configuration for a
new experiment from existing records. It proceeds through guided steps:
selecting the polymer and solvent, aided by a polymer-solvent
compatibility map that summarizes which solvent systems have been
reported for each polymer in the dataset and how frequently; specifying
the needle and collector type from the configurations found in the
matching experiments; and setting a target fiber-diameter mean and
tolerance, shown against the observed diameter distribution of the
matching records. The tool then restricts the records to those matching
the chosen materials, device and target-diameter window, reports the
observed distribution (median and range) of each process parameter across
that subset, and proposes a candidate configuration as the per-parameter
median of the matching records. Crucially, the suggestion is obtained by
retrieving and aggregating existing experimental records, not by inverting
or optimizing a predictive model; the interface describes it as a
candidate configuration derived from existing records and states plainly
that it is not a validated or optimal result.

\begin{figure}[htbp]
  \centering
  \begin{subfigure}[t]{\linewidth}
    \centering
    \includegraphics[width=\linewidth]{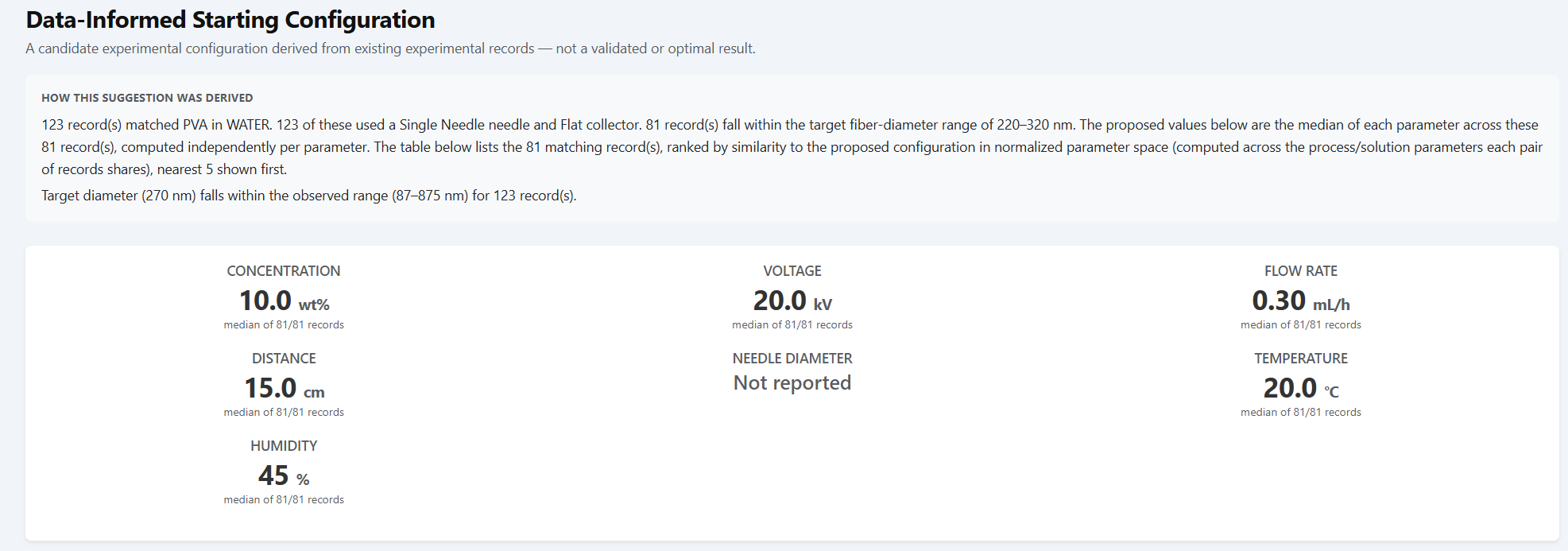}
    \caption{}
    \label{fig:vlab_recipe_a}
  \end{subfigure}
  \\[1ex]
  \begin{subfigure}[t]{\linewidth}
    \centering
    \includegraphics[width=\linewidth]{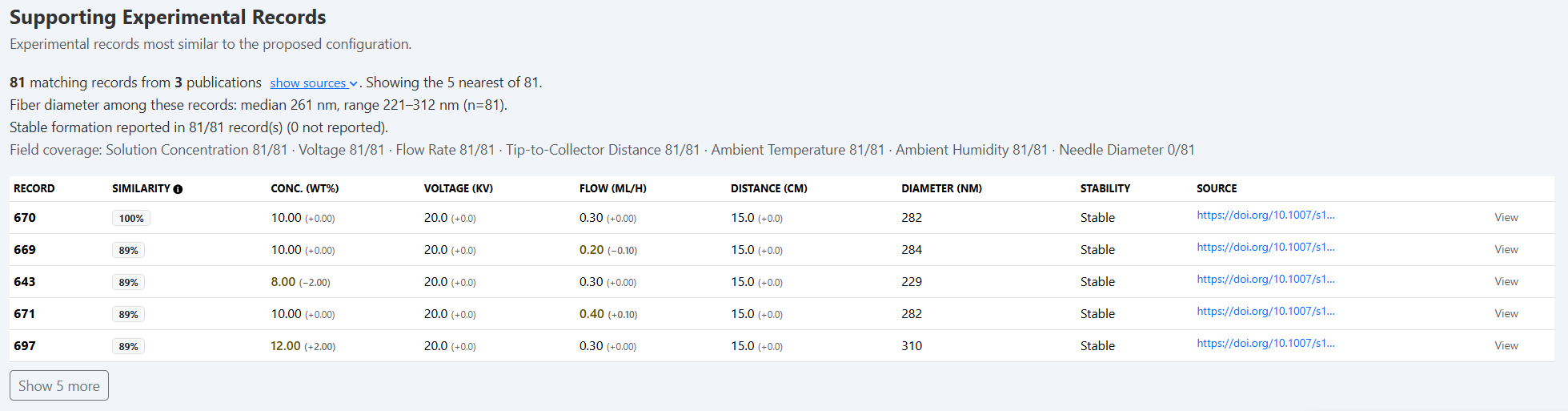}
    \caption{}
    \label{fig:vlab_recipe_b}
  \end{subfigure}
  \caption{Recipe Builder: assembling a data-informed starting
  configuration from existing records, for PVA in water on a
  single-needle, flat-collector setup with a target fiber diameter of
  $270 \pm 50$~nm ($220$-$320$~nm). (a)~Proposed configuration, given as
  the per-parameter median of the 81 matching records that fall within the
  target-diameter window (123 records matched the materials and device; 81
  within the diameter window). The interface labels it a candidate derived
  from existing records and "not a validated or optimal result".
  (b)~Supporting experimental records, ranked by similarity to the
  proposed configuration, showing their measured parameters and fiber
  diameters, the per-field coverage of the matching set (needle diameter
  $0/81$) and links to the source publications, so that each suggested
  value is traceable to the experiments that support it.}
  \label{fig:vlab_recipe}
\end{figure}

Each suggestion is evidence-linked. A supporting-records panel lists the
experimental records most similar to the proposed configuration, ranked
by similarity in normalized parameter space, with their measured process
parameters and fiber diameters, the per-field coverage of the matching
set, and links to the source publications, so that every suggested value
can be traced back to the experiments that support it
(Fig.~\ref{fig:vlab_recipe}). In the example shown, a request for PVA in
water on a single-needle, flat-collector setup with a target fiber
diameter of $270 \pm 50$~nm matches 123 records, of which 81 fall within
the target-diameter window; the proposed configuration
($10$~wt\% concentration, $20$~kV, $0.30$~mL\,h$^{-1}$, $15$~cm
tip-to-collector distance) is the median of those 81 records, which derive
from three publications, and needle diameter is left unreported because it
is absent from all of them ($0/81$ coverage). The resulting configuration
is a starting point grounded in prior evidence and intended for
experimental validation; it is neither a prediction of the fiber diameter
that will result nor an optimized recipe.

\clearpage

\section{Discussion and Outlook} 
Electrospinning-Data.org addresses a persistent limitation in electrospinning research: experimental observations are generated across independent laboratories, material systems, and equipment configurations, but are commonly reported in heterogeneous forms that make their collective reuse difficult. The infrastructure presented here provides a domain-specific mechanism for bringing these observations into a common computational environment in which experimental records can be contributed, harmonized, validated, curated, discovered, and retrieved, and, through the Virtual Lab, analyzed directly. These functions are unified by a shared representation of the electrospinning experiment and a governed data lifecycle, so that contribution, curation, and analysis act on the same records under consistent semantics.
 
This represents an important extension of our previous Cogni-e-SpinDB~1.0 dataset \cite{mahdian_cogni-e-spin_2025, mahdian_cogni-e-spindb_2026}, which established a curated, machine-readable reference of 809 complete experimental records but, like any static release, represents a defined state of the available data. Electrospinning-Data.org extends this model into a continuously evolving infrastructure: new observations can enter through structured web submission, batch ingestion, or programmatic interfaces, undergo harmonization and validation, and, following expert moderation, become part of the live resource. Periodic releases subsequently capture defined states of this evolving dataset as immutable, DOI-bearing archives. The live dataset and archival releases serve complementary purposes: the former supports continued accumulation, discovery, and reuse, whereas the latter provides stable reference points for citation and reproducible downstream analysis.
 
A central design consideration is that aggregation alone does not make heterogeneous experimental observations computationally reusable. Electrospinning experiments combine material composition, solution properties, process parameters, equipment configuration, environmental conditions, and fiber outcomes, while individual publications frequently report only subsets of these variables using different terminology and units. The platform addresses this heterogeneity through a common domain specific schema, canonical units, structured representations of variable experimental configurations \cite{mahdian_electrospinning-dataorg_2026-1}, and controlled morphology descriptors \cite{mahdian_cogni-emcv_2026}, while retaining source and contributor provenance alongside the observations. Records originating from different publications can therefore be queried through the same interfaces and assembled into working datasets without repeating the fundamental harmonization steps for every analysis.
 
An important distinction of Electrospinning-Data.org is its \textit{failure-aware} representation of experimental data. Because stable fibers and desirable morphologies are more likely to be reported, reusable datasets tend to be biased toward successful conditions, yet unstable and unsuccessful experiments delineate the boundaries of feasible operating regions and are therefore scientifically valuable. Rather than requiring every record to represent a success, the platform retains fiber-formation stability, morphological defects, and associated conditions alongside successful outcomes, and records the specific mode of instability through a controlled descriptor set. Accumulating stable, unstable, and unsuccessful experiments in the same representation creates a basis for future analyses of process boundaries, including feasibility mapping, discrimination among instability mechanisms, and experimental designs that account for the risk of unsuccessful processing. The present dataset is not yet dense enough to map these boundaries comprehensively; the contribution is the standardized mechanism through which failure information, when contributed, is preserved and accumulated rather than discarded.

The platform further couples the curated data resource to computational tools through the Virtual Lab, reducing the separation between data discovery and analysis: a researcher can construct a working dataset, inspect its distributions and missingness, examine associations and candidate anomalous observations, train supervised models, retain trained models for reuse, and investigate model behavior through explainability tools within the same environment. The shared working-dataset context matters because analytical outputs remain associated with an explicitly selected subset of the underlying resource rather than an implicitly prepared local copy. The analytical examples presented here should nevertheless be interpreted as demonstrations of the infrastructure rather than validated models of the electrospinning process. Predictive performance remains conditional on the selected material system, experimental records, variables, preprocessing decisions, and evaluation strategy, and models derived from literature-aggregated data should not be assumed to generalize to unrepresented materials, laboratories, or processing configurations.

Beyond lowering the effort of reuse, the Virtual Lab is intended to make the platform immediately useful to individual researchers rather than valuable only in aggregate. A recurring difficulty for community data infrastructures is that the cost of contributing is borne by the contributor while the benefit is diffuse and accrues largely to others, which weakens the incentive to participate. By situating dataset construction, exploratory analysis, modeling, and interpretation directly on the curated data, the Virtual Lab returns value at the point of use: a contributor can analyze their own records, alone or against the aggregated corpus, without assembling a local pipeline, and so obtains a self-contained analytical benefit from engaging with the platform independently of how much the wider community has grown, aligning individual incentive with the collective goal of accumulating reusable data.

The Recipe Builder provides a complementary form of computational reuse that does not depend on fitting a predictive model. Given specified criteria, it retrieves relevant experimental observations, ranks them by similarity, and derives candidate parameter values from the supporting records while exposing the evidence from which those values were obtained. This approach is particularly appropriate at the current scale of the resource, where coverage of the parameter space remains heterogeneous and may not support generalizable predictive models for every material system. The Recipe Builder should therefore be interpreted as an evidence-based parameter-exploration tool rather than an optimizer, inverse-design system, or substitute for experimental validation; its principal value is traceability, since proposed parameters remain connected to experimental observations rather than presented as unsupported recommendations.

As the dataset grows in size and coverage, a natural extension is to complement the evidence-based Recipe Builder with data-driven optimization and inverse design. Where the Recipe Builder retrieves and aggregates existing records, an inverse-design workflow would instead fit a predictive surrogate to the curated data and search the parameter space for configurations expected to yield a target outcome. We have previously demonstrated such an approach for targeted diameter control of electrospun nanofibers, coupling a machine-learning surrogate with optimization to propose process parameters for a desired fiber diameter \cite{mahdian_surrogate-based_2026}. Integrating this class of method into the Virtual Lab, operating on curated dataset versions and retaining traceability to the underlying records—would extend the platform from evidence retrieval toward model-based parameter recommendation. Because inverse design relies on models that generalize across the relevant parameter space, its reliability will remain contingent on sufficiently dense and representative coverage, so we regard it as a capability to be introduced as the resource matures rather than one warranted at the current scale.

The combination of web access, a public REST API, and a Python client allows the resource to operate beyond its own graphical interface. Interactive access supports researchers who browse, filter, and download records directly, whereas the programmatic interfaces allow equivalent queries to be incorporated into scripts, external software, and computational workflows over the same structured resource, so a subset explored interactively can be retrieved programmatically. Authenticated write access additionally establishes a route through which external software and data-producing systems can contribute records using the same schema, providing a foundation for future integration with laboratory data acquisition and instrumentation, although automated end-to-end instrument integration is beyond the scope demonstrated here.
 
The infrastructure operationalizes the FAIR principles at several levels. Experimental records are findable through structured, searchable metadata, and versioned releases receive persistent identifiers through external archival deposition. Interoperability is supported by the common experimental schema, canonical units, machine readable representations, controlled morphology terminology, and the programmatic interfaces described above. Reusability is supported through source and contributor provenance, explicit licensing, and immutable dataset releases that allow downstream studies to reference a defined state of the resource. The continuously evolving platform and the DOI-bearing archival releases are thus complementary components of the FAIR strategy: one permits the scientific resource to evolve, while the other preserves the stable states required for citation and reproducibility.
 
Several limitations remain. First, much of the current resource originates from the published literature and inherits its reporting limitations: parameter coverage is uneven, with environmental conditions and some solution properties reported less consistently than core process variables, and records are distributed unevenly across polymers, solvents, and experimental configurations. Missing information generally cannot be reconstructed reliably when it was not reported by the original investigators. Published literature may also preferentially represent successful experiments, so the failure-aware schema does not by itself eliminate success-reporting bias; realizing its full value will require prospective contributions in which researchers deliberately preserve unsuccessful and unstable experiments alongside successful ones.
 
Second, community contribution remains at an early stage. The long-term value of a continuously evolving infrastructure depends on participation beyond the initial data curation effort and on the addition of independently generated observations. Expert moderation, while important for data quality, introduces a human component that may become a bottleneck as submission volume increases. Automated validation can identify structural inconsistencies, missing required information, dimensional errors, and violations of basic physical constraints, but scientific plausibility cannot be reduced entirely to deterministic rules, so future development will require an appropriate balance between increasingly capable automated validation and expert judgement.

A complementary mechanism, planned as future work, is a controlled embargo option for contributed data. At present the lifecycle moves approved records directly into the public dataset, which may deter researchers from contributing observations they have not yet published or fully analyzed. An embargo would allow a contributor to deposit records and retain private access to them for a defined period—during which the data can be curated and analyzed within the Virtual Lab, before they are released into the public resource, optionally on a fixed schedule or upon publication of the associated study. Such a mechanism would let researchers derive immediate analytical value from their own data while it remains private, and would lower the barrier to early deposition by decoupling the act of contributing from immediate public disclosure, thereby encouraging the prospective, systematically preserved contributions on which the long-term value of the infrastructure depends.

Morphology representation presents a further challenge. The controlled taxonomy provides a common vocabulary for describing fiber characteristics and reduces the ambiguity of unstructured descriptions, but electrospun fiber morphology is diverse and the representation will need to evolve as new structures and descriptors are encountered. Although image evidence can be associated with records, automated extraction of quantitative morphological information from microscopy images is not currently integrated into the data lifecycle. Linking the controlled morphology representation with reproducible image-analysis methods is an important opportunity for future development and could ultimately enable images, structured morphology descriptors, and quantitative measurements to be treated as complementary representations of the same experimental outcome.
 
Long-term sustainability and interoperability are also important for a continuously operated infrastructure. The separation between the live operational database and DOI-bearing archival releases reduces dependence on the continued availability of the web service for access to previously released dataset states. Nevertheless, maintaining the interactive platform, moderation workflow, APIs, computational services, and evolving data model requires continued technical and community support. Future development should therefore prioritize capabilities that strengthen the underlying scientific resource rather than simply increasing the number of software features.
 
Electrospinning-Data.org consequently represents a transition from a single curated electrospinning dataset toward a persistent, community-oriented and failure-aware data infrastructure. Its current significance lies not in providing a complete representation of the electrospinning parameter space, which remains beyond the scale and coverage of the available data, but in establishing the mechanisms through which that representation can improve over time: a shared experimental schema capable of representing successful and unsuccessful outcomes, controlled terminology, provenance, validation and expert moderation, open and programmatic access, versioned archival releases, and computational tools operating directly on the curated resource. As the number and diversity of contributed experiments increase, particularly through the prospective inclusion of unsuccessful and unstable experiments, the infrastructure can provide an increasingly informative basis for cross-study comparison, reproducible data-driven analysis, identification of feasible and failure-prone processing regimes, and systematic investigation of electrospinning process-structure-property relationships.

\clearpage

\section*{Funding}
This study was funded by the Estonian Research Council (grant PSG897). The funder had no role in the study design, data collection, data analysis, interpretation of results, or writing of the manuscript.

\section*{Competing interests}
The authors declare no competing interests.

\section*{Data availability}

The platform and experimental data are publicly available through Electrospinning-Data.org at https://electrospinning-data.org/. Versioned dataset releases are first issued through the platform and subsequently archived on Zenodo \cite{mahdian_cogni-e-spin_2025-1}.

\section*{Code availability}
The source code underlying the Electrospinning-Data.org platform is openly available at \url{https://github.com/taltechloc/electrospinning-data.org}.

\section*{Author contributions}
M.M. conceived the study, developed the methodology, designed the data infrastructure, implemented the software platform, curated the dataset, performed the formal analysis, and wrote the original manuscript draft. T.P. and F.E. supervised the project, provided resources, and reviewed and edited the manuscript. All authors approved the final version of the manuscript.

\clearpage
\bibliography{ref}

\end{document}